\documentclass[11pt]{amsart}

\newtheorem{Theorem}{Theorem}

\newtheorem{Definition}{Definition}

\newtheorem{Lemma}{Lemma}

\begin{document}

\title[Quasi-Ballistic Dynamics]{Quantum Hamiltonians with
Quasi-Ballistic Dynamics and
Point Spectrum}
\author{C\'{e}sar R. de Oliveira$^1$}
\thanks{$^{(1)}$ Corresponding  author and partially supported by CNPq (Brazil).}
\address{Departamento de Matem\'{a}tica -- UFSCar, S\~{a}o Carlos, SP,
13560-970 Brazil}
\email{oliveira@dm.ufscar.br}
\author{Roberto A. Prado$^2$}
\thanks{$^{(2)}$ Supported by CAPES (Brazil).}
\address{Departamento de Matem\'{a}tica -- UFSCar, S\~{a}o Carlos, SP,
13560-970 Brazil\\}
\email{rap@dm.ufscar.br}
\subjclass{81Q10 (11B,47B99)}

\begin{abstract} Consider the family of Schr\"o\-dinger operators (and
also its Dirac version) on
$\ell^2(\mathbb{Z})$ or $\ell^2(\mathbb{N})$
\[ H^W_{\omega,S}=\Delta + \lambda F(S^n\omega) + W , \quad
\omega\in\Omega ,
\]
 where $S$ is a transformation on (compact metric)
$\Omega$, $F$ a real Lipschitz function and $W$ a (sufficiently fast)
power-decaying perturbation. Under
certain conditions it is shown that $H^W_{\omega,S}$ presents
quasi-ballistic dynamics for
$\omega$ in a dense $G_{\delta}$ set. Applications include potentials
generated by rotations of the torus
with analytic condition on $F$, doubling map, Axiom~A dynamical systems
and the Anderson model. If $W$
is a rank one perturbation, examples of $H^W_{\omega,S}$ with
quasi-ballistic dynamics and
 point spectrum are also presented.
\end{abstract}
\maketitle

\section{Introduction} Quantum Hamiltonians, i.e., Schr\"odinger and
Dirac, with potentials along
dynamical systems is a very interesting subject that has been
considered in the mathematics and physics
literature, mainly one-dimensional discrete versions. Although not
explicitly stated, it is natural to
expect that the more ``chaotic'' the underlining dynamical system, the
more singular the corresponding
spectrum; the extreme cases could be represented by periodic potentials
on one hand, which impose
absolutely continuous spectrum and ballistic dynamics (see
Definition~\ref{ballDef}), and random
potentials on the other hand, that lead to point spectrum and absence
of transport (bounded moments of
the position operator). We mention the
papers~\cite{CadO,DLS,DST,DT,dOP1,dOP2,GKT,JSBS,T} for references
and additional comments on important recent results on quantum dynamics
for Dirac and Schr\"o\-dinger
operators.

Exceptions of the above picture are known, since there are examples of
one-dimensional quantum models
with pure point spectrum and transport. Here we refer to the random
dimer model~\cite{JSBS} for the
Schr\"odinger case and the random Bernoulli Dirac
operator~\cite{dOP1,dOP2} (with no potential
correlation). The first example of (Schr\"odinger) operators with such
``unexpected behavior'' has
appeared in~\cite{dRJLS}, Appendix~2, what the authors have called ``A
Pathological Example;'' the
potential was the almost-Mathieu (see Application~\ref{AlmostMathieu}
ahead), which is built along
irrational rotations of the circle, with a combination of suitable
rational approximations for the
rotation angle and a rank one perturbation.

Rotations of the circle are by far the most considered dynamical
systems to generate quantum potentials
\cite{J,BJ,Pu}; their finite-valued versions \cite{BIST,DP,DL},
together with substitution dynamical system
 potentials (see~\cite{Le,LdO} and references therein) are
mathematical models of
one-dimensional quasi-crystals with predominance of singular continuous
spectrum. These dynamical systems
are not ``chaotic,'' which could be characterized by positive
entropy~\cite{KH} or via a more dynamical
definition gathered in~\cite{Dev}; the paradigms of chaotic systems are
the Anosov
and, more generally,  Axiom~A
systems.

Since chaotic motion mimics randomness, it is natural to conjecture
that for quantum operators with
suitable potentials built along Axiom~A (and other chaotic) systems there
is a
predominance of point spectrum and
absence of transport. A small step in this direction are the results
of~\cite{BoS} about Anderson
localization for potentials related to the doubling map $\theta\mapsto
2\theta$ on the circle and also
hyperbolic toral automorphisms---both systems have positive entropy.

The main goal of this paper is to have a close inspection on the
construction of the above mentioned
``unexpected example'' in~\cite{dRJLS}, together with the related
analysis in~\cite{GKT}, in order to get
a different view of them and so provide new examples of quantum operators
with quasi-ballistic dynamics,
some of them with pure point spectrum. In spite of the above
conjecture, as applications we
can prove that for a generic (i.e., dense
$G_\delta$) set of initial conditions of Axiom~A systems, as well as of
chaotic dynamical systems as
defined in Devaney~\cite{Dev},
 the associated quantum operators present quasi-ballistic dynamics. We
will also have
something to say
 about the random Anderson model, that is, there is a dense
$G_\delta$ set of initial conditions so that the quantum operators
present quasi-ballistic dynamics; see
Section~\ref{ApplicationsSection} for details and
 other examples. The applications are the principal contributions of
this paper. From now on we shall
formulate more precisely the context we work at.

 Let $(\Omega,d)$ be a compact metric space. Consider the family of
bounded Schr\"o\-dinger operators
$H^W_{\omega,S}$ given by
\begin{equation} \label{model} (H^W_{\omega,S}\psi)(n)=(\Delta\psi)(n) +
\lambda F(S^n\omega)\psi(n) + W(n)\psi(n) ,
\quad
\omega\in\Omega ,
\end{equation} acting on $\psi\in\ell^2(\mathbb{N})$ (with a Dirichlet,
or any other, boundary condition)
or the whole lattice case $\ell^2(\mathbb{Z})$, where the Laplacian
$\Delta$ is the finite difference operator
\[ (\Delta\psi)(n)=\psi(n+1)+\psi(n-1),
\]
 $S$ is a transformation on
$\Omega$ (invertible in the whole lattice case),
$F:\Omega\to \mathbb{R}$ satisfies a Lipschitz condition, i.e., there
exists
$L>0$ such that
\begin{equation} \label{Lipschitz} |F(\theta)-F(\omega)| \leq L
d(\theta,\omega),
\quad \forall \
\theta,\omega\in\Omega,
\end{equation} and, for some $\eta>0$ and $0<\tilde{C}<\infty$, the
perturbation
$W$ satisfies
\begin{equation} \label{perturbation} |W(n)| \leq
\tilde{C}(1+|n|)^{-1-\eta}, \quad \forall \ n\in\mathbb{Z}.
\end{equation} The coupling constant $\lambda$ is a positive real
number. Throughout $W$ is supposed to
satisfy~(\ref{perturbation}). We shall denote by $\ell$ the Lebesgue
measure (normalized, when necessary)
and by
$\sigma(H)$ the spectrum of a self-adjoint operator~$H$.

 We are interested in situations where nontrivial quantum transport for
systems governed by
the above Hamiltonians can be established. To this end consider
the time averaged moments of
order $p>0$ associated to the initial state
$\delta_1$ (a member of the canonical basis of $\ell^2$), defined by
\begin{equation} \label{dynamic moments}
M^W_{\omega,S}(p,T):=\frac{2}{T}\int_0^{\infty}{e^{-2t/T}}
\sum_n (1+n^2)^{p/2} \ |\langle \delta_n, e^{-itH^W_{\omega,S}}\delta_1
\rangle|^2\ dt.
\end{equation} The presence of quantum transport will be probed through
the upper diffusion exponents
\begin{equation} \label{Transp.Exponents}
\beta_{\omega,S,W}^+(p):=\limsup_{T \to \infty}\frac{\log
M^W_{\omega,S}(p,T)}{p\
\log T}\ .
\end{equation} The lower diffusion exponents will be denoted by
\begin{equation} \label{Transp.ExponentsL}
\beta_{\omega,S,W}^-(p):=\liminf_{T \to \infty}\frac{\log
M^W_{\omega,S}(p,T)}{p\
\log T}\ .
\end{equation}

\

\begin{Definition}\label{ballDef} If $\beta_{\omega,S,W}^-(p)=1$ for all
$p>0$, the operator
$H^W_{\omega,S}$ is said to present ballistic dynamics. If
$\beta_{\omega,S,W}^+(p)=1$ for all $p>0$, the operator
$H^W_{\omega,S}$ is said to present
quasi-ballistic dynamics.
\end{Definition}

Although point spectrum has been associated with localized dynamics, as
already mentioned the first
example of a Schr\"o\-dinger operator with quasi-ballistic dynamics and
point spectrum was the half
lattice almost Mathieu operator under rank one
perturbation~\cite{dRJLS}. The random dimer
model~\cite{DBG} and the Bernoulli-Dirac model~\cite{dOP1,dOP2} (zero
mass case) are other examples of
operators with nontrivial quantum transport (due to existence of
critical energies) and pure point
spectrum. In~\cite{GKT} a new method was developed to obtain dynamical
lower bounds with application for
random decaying potentials.

Here we are confined to quasi-ballistic transport; the ideas
in~\cite{dRJLS} for the almost Mathieu
operator, and then revisited in~\cite{GKT}, is presented from a rather
different viewpoint in order to
provide new examples of quantum operators with quasi-ballistic
dynamics, some of them also with pure point
spectrum (see ahead).

The abstract result we shall present can be summarized as (see
Theorem~\ref{Transport} for a precise
statement): If there exists a dense set of initial conditions in
$\Omega$ for which the transfer matrices are bounded from above in
energy intervals with positive
Lebesgue measure, and if the iterations of
$S$ satisfies a suitable continuity-like condition, then one obtains a
dense
$G_{\delta}$ set
$\tilde{\Omega}\subset\Omega$ such that for any
$\omega\in\tilde{\Omega}$,
$H^W_{\omega,S}$ defined by~(\ref{model}) presents quasi-ballistic
dynamics. With respect to the spectral
type, we shall highlight a known result (see Theorem~\ref{Point
Spectrum}) that will be used in some
applications: If the Lyapunov exponent corresponding the
$H_{\omega,S}^{W=0}$ is strictly positive for energies in the spectrum,
then under the rank one
perturbation $W=\kappa \langle\delta_1,\cdot\rangle\delta_1$ the
Schr\"odinger operator
$H^W_{\omega,S}$ on~$\ell^2(\mathbb{N})$ has pure point spectrum for
a.e.\ $\omega$ (with respect to an
ergodic measure) and a.e.\
$\kappa$ (with respect to Lebesgue $\ell$ measure). There is a
restricted version for the whole lattice
case. That result will be so important for some applications here that a
sketch of its proof will be
provided. We shall apply the
abstract result to several types of potential
$V_{\omega}(n)=F(S^n\omega)$ (see Section~\ref{ApplicationsSection}):
Rotations of
$\mathbb{S}^1$ and of the torus with analytic condition on $F$,
doubling map, Anderson model, Anosov and
Axiom~A, and chaotic dynamical systems. For the particular case of
incommensurate rotations of the torus
under rank one perturbations (see Subsections~\ref{appl.1}
and~\ref{appl.2}), besides quasi-ballistic
dynamics, it is also found the concomitant presence of pure point spectrum.

This paper is organized as follows: In Section~\ref{ResultSection} the
results about quasi-ballistic
dynamics (Theorem~\ref{Transport}) and point spectrum
(Theorem~\ref{Point Spectrum}) for the model
(\ref{model}) are presented, whose proofs appear in
Section~\ref{ProofsSection}. In
Section~\ref{PreliminarySection} some preliminary results used in those
proofs are collected.
Section~\ref{ApplicationsSection} is devoted to applications. In
Section~\ref{DiracSection} the
adaptation of the results for the discrete Dirac model is briefly
mentioned.

\

\section{Abstract Results}
\label{ResultSection} In this section we will present our result about
quasi-ballistic dynamics
(Theorem~\ref{Transport}) for the operators
$H^W_{\omega,S}$ defined by (\ref{model}) and also a spectral result
(Theorem~\ref{Point Spectrum}) that
will be used in some applications. First of all, we recall the notion
of transfer matrices. These matrices
$\Phi^W_{\omega,S}$ are uniquely defined by the condition that
\[ \left( \begin{array}{c}\psi(n+1) \\ \psi(n) \\
\end{array}\right) = \Phi^W_{\omega,S}(E,n,0) \left(
\begin{array}{c}\psi(1) \\ \psi(0) \\
\end{array}
\right) \] for every solution $\psi$ of the eigenvalue equation
\[ H^W_{\omega,S}\psi=E\psi.
\] Hence,
\[ \Phi^W_{\omega,S}(E,n,0)=
\left\{ \begin{array}{ccc} T^W_{\omega,S}(E,n)\ldots
T^W_{\omega,S}(E,1) &
\quad n\geq 1, \\\\ Id &
\quad n=0, \\\\
\big(T^W_{\omega,S}(E,n+1)\big)^{-1}\ldots
\big(T^W_{\omega,S}(E,0)\big)^{-1} & \ \ \quad n\leq -1,
\end{array} \right.\] where
\[ T^W_{\omega,S}(E,k)=\left(
\begin{array}{cc} E-\lambda F(S^k\omega)-W(k) & -1 \\ \\ 1 & 0
\end{array} \right).\]

\

Now we are in position to state the main abstract result of this paper.

\

\begin{Theorem}\label{Transport} Let $H^W_{\omega,S}$ be the operator
defined by (\ref{model}) on
$\ell^2(\mathbb{N})$ with $S$ and $W$ (as in~(\ref{perturbation}))
fixed. Suppose that there exists a
dense set
$A$ of initial conditions in
$\Omega$ such that, for each
$\omega\in A$, there are a closed interval $J^{\omega}_S
\subset\sigma(H^0_{\omega,S})$ (i.e., the spectrum in the case $W\equiv
0$) with $\ell(J^{\omega}_S)>0$
and
$0<C_{\omega}(S)<\infty$ so that
\begin{equation}\label{rel2Theorem1}
\big \|\Phi^{0}_{\omega,S}(E,n,0)\big \|\leq C_{\omega}(S), \quad
\forall\ E\in J^{\omega}_S \ \ \mbox{and} \ \ \forall\ n\in\mathbb{N}.
\end{equation}
 Assume that there are $0<C<\infty$ and a nonnegative function
$h_S:\mathbb{N} \to \mathbb{R}$
satisfying
\[ d(S^n\theta,S^n\omega)\leq C d(\theta,\omega) h_S(n), \quad
\forall\ \theta,\omega\in\Omega \ \ \mbox{and} \ \ \forall\
n\in\mathbb{N}.
\]
 Then there exists a dense
$G_{\delta}$ set
$\tilde{\Omega}\subset\Omega$ such that, for each
$\omega\in\tilde{\Omega}$, the operator $H^W_{\omega,S}$ presents
quasi-ballistic dynamics.
\end{Theorem}

\

 Let $\nu$ be an ergodic probability measure on $\Omega$ with respect
to~$S$. By Furstenberg and Kesten
Theorem~\cite{BLa}, for
$\nu-a.s.$ $\omega$ the Lyapunov exponent
\[\Gamma^W_{S}(E)=\lim_{n\to\infty} \frac{1}{|n|} \log
\big \| \Phi^W_{\omega,S}(E,n,0)\big \|
\] exists and is independent of $\omega$. The next result is a
consequence of the Simon-Wolff
criterion~\cite{SiW} (see Lemma~\ref{pointspecLemma} ahead). Recall
that the cyclic subspace generated by
$\phi\in \ell^2$ for a self-adjoint operator~$H$ is the closure of
$\{(H-z)^{-1}\phi: z\in\mathbb{C}\}$; the vector $\phi$ is cyclic for
$H$ if such subspace is the
whole~$\ell^2$.

\

\begin{Theorem}[\cite{Si}]\label{Point Spectrum} Let $H^W_{\omega,S}$
be the operator defined by
(\ref{model}) on
$\ell^2(\mathbb{Z})$ under rank one perturbations $W=~\kappa\langle
\delta_1,\cdot\rangle
\delta_1$, \ $\kappa \in\mathbb{R}.$ Fix an interval $[a,b]$. If
$\Gamma^0_{S}(E)>0$ for $ \ell-a.s. \ E\in [a,b]$, then restricted to
the cyclic subspace generated by
$\delta_1$, the operator $H^W_{\omega,S}$ has pure point spectrum in
$[a,b]$ for $\ell-a.s.\ \kappa$ and $\nu-a.s.\
\omega$.
\end{Theorem}

\

\noindent \textit{Remarks.}
\begin{itemize}
\item[i)] Theorem~\ref{Transport} can be readily adapted to the whole
lattice case.
\item[ii)] The set $\tilde \Omega$ in Theorem~\ref{Transport} does not
depend on the perturbation $W$
(including $W\equiv0$).
\item[iii)] Although both theorems above have half and whole lattice
versions, the proofs of such versions
are quite similar; so Theorem~\ref{Transport} will be proven for the
half lattice case while
Theorem~\ref{Point Spectrum} for the whole lattice one.
\item[iv)] Since in the half lattice case the vector $\delta_1$ is
cyclic for
$H^0_{\omega,S}$, then in this case the conclusions of
Theorem~\ref{Point Spectrum} hold
on~$\ell^2(\mathbb{N})$.
\item[v)] In this paper (and perhaps in most future applications) the
set $A$ in
Theorem~\ref{Transport} is composed of periodic orbits of the map~$S$.
\end{itemize}

\

\section{Preliminaries}
\label{PreliminarySection} In this section we collect some results that
will be used in the proofs of
Theorems~\ref{Transport} and~\ref{Point Spectrum}. Most of them are
know results whose proofs are easily
found in the references. Denote by
$\mu^W_{\omega,S}$ the spectral measure associated to the pair
$(H^W_{\omega,S},\delta_1)$ and introduce the ``local spectral moments''
\cite{GKT}
\begin{equation}\label{defK}
K_{\mu^W_{\omega,S}}(q,\epsilon):=\frac{1}{\epsilon}
\int_{\mathbb{R}}
\left(\mu^W_{\omega,S}(x-\epsilon,x+\epsilon)\right)^q dx
\end{equation}
 defined for $q>0$ and $\epsilon>0$. A key point for the proof of
Theorem~\ref{Transport} will be the
following lower bound for the diffusion exponents
$\beta_{\omega,S,W}^+(p)$.

\

\begin{Lemma}\label{Liminfexponents} For all $p>0$ and $q=(1+p)^{-1}$,
one has
\[
\beta_{\omega,S,W}^+(p)\geq \limsup_{\epsilon\to 0}
\frac{\log K_{\mu^W_{\omega,S}}(q,\epsilon)}{(q-1)\log\epsilon}\ .
\]
\end{Lemma}

\

 The proof of Lemma~\ref{Liminfexponents} follows directly from
Theorem~2.1 of~\cite{BGT1} and Lemmas~2.1
and~2.3 of~\cite{BGT2}. The next result converts an upper bound on the
norm of transfer matrices into a
lower bound on the spectral measure; for its proof see Proposition~2.1
of~\cite{GKT}.

\

\begin{Lemma} \label{specmeas,transMatri} Let $H^W_{\omega,S}$ be the
operator defined by~(\ref{model}) on
$\ell^2(\mathbb{N})$ and let $I$ be a compact interval. There exist a
universal constant
$C_1$ and, for all $M>0$ and $\tau >0$, a constant $C_2=C_2(I,M,\tau)$
such that for all
$\epsilon\in (0,1)$ and all $x\in I$, one has
\[
\mu^W_{\omega,S}(x-\epsilon,x+\epsilon)\geq C_1
\int_{x-\frac{\epsilon}{2}}^{x+\frac{\epsilon}{2}}\frac{dE}{\big
\| \Phi^W_{\omega,S}(E,N,0)\big \|^2}\ -\ C_2\epsilon^M \ ,
\] with
$N=[\epsilon^{-1-\tau}]$ (integer part).
\end{Lemma}

\

 In order to establish relations between the transfer matrices with
different initial conditions, the
next result will be used.

\

\begin{Lemma} \label{Form.Var.c.i.} Let $E\in\mathbb{R}$, $N>0$ and set
\[ L_S^{\omega}(N):=\sup_{1\leq n \leq N} \big
\| \Phi^{W}_{\omega,S}(E,n,0)\big \|\ .
\] Then, for $ 1\leq n \leq N$ and $\theta\in\Omega$,
\[
\big \| \Phi^W_{\theta,S}(E,n,0)\big \|\leq L_S^{\omega}(N)
e^{L_S^{\omega}(N)
\lambda |F(S^n\theta) - F(S^n\omega)| n}.
\]
\end{Lemma}
\begin{proof} An inductive argument shows that, for $\theta, \omega \in
\Omega$ and $n\ge1$, one can write the identity
\[
\Phi^W_{\theta,S}(E,n,0)= \Phi^{W}_{\omega,S}(E,n,0)+ \lambda
\sum_{j=1}^{n} \Phi^{W}_{\omega,S}(E,n,j)\ B_S^{\theta,\omega}(n)\
\Phi^W_{\theta,S}(E,j,1)\ ,
\] where
\[B_S^{\theta,\omega}(n) =\left(
\begin{array}{cc} F(S^n\omega) - F(S^n\theta) & 0 \\ \\ 0 & 0
\end{array} \right).\] By iteration, using the fact that $\big \|
\Phi^{W}_{\omega,S}(E,n,0)\big \|\leq L_S^{\omega}(N)$ for all
$1\leq n \leq N$, one obtains
\begin{eqnarray*}
\big \|\Phi^W_{\theta,S}(E,n,0)\big \| &\leq& L_S^{\omega}(N)
\left[1+\lambda |F(S^n\theta) - F(S^n\omega)|\:
L_S^{\omega}(N)\right]^{n-1} \\ &\leq& L_S^{\omega}(N)
e^{L_S^{\omega}(N) \lambda |F(S^n\theta) - F(S^n\omega)| n},
\end{eqnarray*} for $1\leq n \leq N$.
\end{proof}

\

 Now we describe two results that will be used in the proof of
Theorem~\ref{Point Spectrum}. Details will
be presented only for the whole lattice case. Consider the function
\[ G_{\theta,S}(E)=\int \frac{d\mu^0_{\theta,S}(x)}{(E-x)^2}\
\] which is defined for
$E\in (-\infty,\infty)$ and takes values in
$(0,\infty]$. The first result relates $G_{\theta,S}(E)$ with the
solutions of the eigenvalue equation
\begin{equation} \label{eigenequation} H_{\theta,S}^{0}\psi=E\psi.
\end{equation} See Theorem~2.4 of~\cite{Si} for its proof.

\

\begin{Lemma}\label{solutionLemma} Let $H_{\theta,S}^{0}$ be the
operator defined by~(\ref{model}) on
$\ell^2(\mathbb{Z})$, with $W\equiv0$. Then one has
$G_{\theta,S}(E)<\infty$ if and only if
\begin{itemize}
\item[(i)] E is not an eigenvalue of $H_{\theta,S}^{0}$;
\item[(ii)] One of the following holds:
\begin{itemize}
\item[(ii.1)] equation~(\ref{eigenequation}) has an $\ell^2$ solution on
$(0,\infty)$ with $\psi(0)=0$;
\item[(ii.2)] (\ref{eigenequation}) has an $\ell^2$ solution on
$(-\infty,0)$ with $\psi(0)=0$;
\item[(ii.3)] (\ref{eigenequation}) has an $\ell^2$ solutions
$\psi_\pm$ on both $(0,\infty)$ and $(-\infty,0)$ with both
$\psi_+(0)\neq0$ and $\psi_-(0)\neq0$.
\end{itemize}
\end{itemize}
\end{Lemma}

\

 Finally we remind of Simon-Wolff criterion~\cite{SiW}:

\

\begin{Lemma}\label{pointspecLemma} Let $H^W_{\omega,S}$ be the
operator defined by~(\ref{model}) on
$\ell^2(\mathbb{Z})$ with $W=\kappa \langle
\delta_1,\cdot\rangle\delta_1$, $\kappa\in\mathbb{R}$. Fix an interval
$[a,b]$. Then the following assertions are equivalent:
\begin{itemize}
\item[(i)] $G_{\omega,S}(E)<\infty$ for $\ell-a.s. \ E\in [a,b]$;
\item[(ii)] restricted to the cyclic subspace generated by
$\delta_1$, the operator $H^W_{\omega,S}$ has only pure point spectrum
in
$[a,b]$ for $\ell-a.s. \ \kappa.$
\end{itemize}
\end{Lemma}

\

\section{Proofs}
\label{ProofsSection} In this section the proofs of
Theorems~\ref{Transport} and ~\ref{Point Spectrum}
are presented. In order to prove Theorem~\ref{Transport}, the following
technical result will be used.

\

\begin{Lemma} \label{q-b.motion} Let $A$ be the set described in
Theorem~\ref{Transport} and fix
$\omega\in A$. Then there exists $\epsilon(\omega,S)>0$ such that for
every
$0<\epsilon<\epsilon(\omega,S)$ it is possible to choose
$\delta(\epsilon,\omega,S)>0$ such that if \
$d(\theta ,\omega)<\delta(\epsilon,\omega,S)$, then for any
$q\in (0,1)$ there exists $0<C_q<\infty$ so that
\[ K_{\mu^W_{\theta,S}}(q,\epsilon)\geq \ C_q \
\frac{\epsilon^{-1+q}}{\log(\epsilon^{-1})}\ .
\]
\end{Lemma}
\begin{proof} For each $\omega\in A$ fixed, there exists a closed
interval
$J_S^{\omega}\subset\sigma(H^0_{\omega,S})$ with
$\ell(J_S^{\omega})\geq L_{\omega}(S)>0$ and
$0<C_{\omega}(S)<\infty$ such that
\[
\big \|\Phi^0_{\omega,S}(E,n,0)\big \|\leq C_{\omega}(S), \quad \
\forall\ E\in J^{\omega}_S \ \ \mbox{and} \ \ \forall\ n\in\mathbb{N}.
\] Since
$W$ satisfies~(\ref{perturbation}),  it is
found that
\begin{equation} \label{limsupTransMatr}
\big \|\Phi^W_{\omega,S}(E,n,0)\big \|^2\leq
\tilde{C}_{\omega}(S), \quad \ \forall\ E\in J^{\omega}_S \ \
\mbox{and} \ \ \forall\ n\in\mathbb{N}.
\end{equation} We  remark that inequality (\ref{limsupTransMatr}) is closely related to discrete
versions of the Levinson's theorem (see, e.g., \cite{JM,Sil} and references there in), but in Theorem~2
of~\cite{DST}  a detailed and ad hoc proof is presented. 

Pick
$\epsilon(\omega,S)>0$ such that if
$\epsilon <\epsilon(\omega,S)$, then for any
$q\in (0,1)$,
\begin{equation} \label{cond.Cresc.}
\max\{\tilde{C}_{\omega}(S),L_{\omega}(S)^{-1}\}\leq
\left(\log(\epsilon^{-1})\right)^{1/(1+q)} .
\end{equation} Now note that, by~(\ref{Lipschitz}) and the hypotheses
of Theorem~\ref{Transport}, one has
\begin{equation} \label{cond.Continuity} |F(S^n\theta) -
F(S^n\omega)|\leq L d(S^n\theta,S^n\omega)
\leq L C d(\theta,\omega)h_S(n) ,
\end{equation} for every $\theta,\omega\in\Omega$ and for all
$n\in\mathbb{N}$. Note that by using the new function $H_S(n):=\max_{0\le j\le n} h_S(j)$, one may assume
that $h_S$ is nondecreasing; this will be done in what follows.

Pick
$\tau >0$. As a consequence
of~(\ref{limsupTransMatr}),~(\ref{cond.Continuity}) and
Lemma~\ref{Form.Var.c.i.}, it is found that for $\omega\in A$ fixed and
for any
$\epsilon <\epsilon(\omega,S)$,
\begin{eqnarray} \label{LimsupTransMatr2}
\big \|\Phi^W_{\theta,S}(E,n,0)\big \|^2 & \leq &
\tilde{C}_{\omega}(S)e^{2\lambda \tilde{C}_{\omega}(S)L C
d(\theta,\omega)h_S([\epsilon^{-1-\tau}])\epsilon^{-1-\tau}} \\ & \leq
& 2\lambda L C
\tilde{C}_{\omega}(S) , \nonumber
\end{eqnarray} for every $E\in J_S^{\omega}$ and for all $1\leq n\leq
[\epsilon^{-1-\tau}]$, where we
required that $d(\theta,\omega)$ is small enough (which determines
$\delta(\epsilon,\omega,S)>0$) so that
\[ 2\lambda \log(\epsilon^{-1})L C
d(\theta,\omega)h_S([\epsilon^{-1-\tau}])\epsilon^{-1-\tau}\leq
\log(2\lambda L C).
\] Thus, by Lemma~\ref{specmeas,transMatri} with $M=2$ and by
(\ref{cond.Cresc.}) and
(\ref{LimsupTransMatr2}), it follows that for
$\epsilon$ small enough,
\begin{eqnarray*}\label{Liminfspecmeas}
\mu^W_{\theta,S}(E-\epsilon,E+\epsilon) & \geq & C_1
\big(2\lambda L C \tilde{C}_{\omega}(S)\big)^{-1}\epsilon - C_2
\epsilon^2 \\ & \geq & C_3\
\frac{\epsilon}{\left(\log(\epsilon^{-1})\right)^{1/(1+q)}}\ ,
\end{eqnarray*} for every $E\in J_S^{\omega}$. Therefore, for any $q\in
(0,1)$ and
$\epsilon < \epsilon(\omega,S)$, it follows from (\ref{defK}),
(\ref{cond.Cresc.}) and the above
inequality that
\[ K_{\mu^W_{\theta,S}}(q,\epsilon)\geq C_q\
\frac{\epsilon^{-1+q}}{\big(\log(\epsilon^{-1})\big)^{q/(1+q)}}\
\ell(J_S^{\omega})\geq C_q\
\frac{\epsilon^{-1+q}}{\log(\epsilon^{-1})}\ .
\]
\end{proof}

\

\noindent \textit{Remarks.} Both Lemma~\ref{q-b.motion} and the proof of Theorem~\ref{Transport} hold
if the logarithm function is replaced by any $g:\mathbb{R}\to\mathbb{R}$ with $\lim_{t\to\infty}g(t)=\infty$ and
$\lim_{t\to\infty}g(t)/t=0$.

\

\begin{proof} {\bf (Theorem~\ref{Transport})} For each
$n\in\mathbb{N}\setminus\{0\}$ define the sets
\[B_n= \bigg\{\theta\in\Omega\ \Big|\ \exists\ \epsilon<\frac{1}{n} :
\ K_{\mu^W_{\theta,S}}(q,\epsilon)\geq \ C_q \
\frac{\epsilon^{-1+q}}{\log(\epsilon^{-1})} \bigg\}. \] Since $A$ is
dense, by Lemma~\ref{q-b.motion}
each of the sets $B_n$ contains a dense open set. Therefore, by Baire
Theorem,
$\bigcap_{n=1}^{\infty}B_n$ contains a dense $G_{\delta}$ set
$\tilde{\Omega}$. Note that for each $\theta\in \tilde{\Omega}$ there
exists a sequence $\epsilon_n
\rightarrow 0$ such that
\[ K_{\mu^W_{\theta,S}}(q,\epsilon_n)\geq \ C_q \
\frac{\epsilon_n^{-1+q}}{\log(\epsilon_n^{-1})}\ ,
\] for any $q\in (0,1).$ Choosing
$q=(1+p)^{-1}$, it follows by Lemma~\ref{Liminfexponents} that for any
$\theta\in\tilde{\Omega}$ and for all $p>0$, \
$\beta_{\theta,S,W}^+(p)=1$, i.e., the operator $H_{\theta,S}^{W}$
presents quasi-ballistic dynamics.
\end{proof}

\

\begin{proof}{\bf (Theorem~\ref{Point Spectrum})} By hypothesis,
$\Gamma^0_{S}(E)>0$ for $\ell-a.s. \ E\in [a,b]$. The Theorem of
Ruelle-Oseledec
\cite{Ru} implies that there exist solutions $\psi_\pm$ of the equation
(\ref{eigenequation}), for
$\ell-a.s. \ E\in[a,b]$, that are $\ell^2$ at $\pm\infty$ (they decay
exponentially). Hence by
Lemma~\ref{solutionLemma}, either
$E$ is an eigenvalue of $H_{\theta,S}^{0}$ or
$G_{\theta,S}(E)<\infty$. Since $H_{\theta,S}^{0}$ has only countably
many eigenvalues, it is possible to
conclude that
$G_{\theta,S}(E)<\infty$ for $\ell-a.s. \ E\in [a,b]$. Therefore, it
follows by
Lemma~\ref{pointspecLemma} that, restricted to the cyclic subspace
generated by
$\delta_1$, the operator $H^W_{\theta,S}$ has pure point spectrum in
$[a,b]$ for $\nu-a.s. \ \theta$ and
$\ell-a.s. \ \kappa$.
\end{proof}

\

\section{Applications}
\label{ApplicationsSection} This section is devoted to applications of
Theorems~\ref{Transport}
and~\ref{Point Spectrum}. Some of them provide examples of quantum
operators with quasi-ballistic
dynamics and point spectrum (pure point in the half lattice case).

\

\subsection{Anosov and Axiom A}
\label{appl.AxiomA} Let $M$ be a differentiable compact manifold.
Recall that a diffeomorphism
$S:M\to M$ satisfies the Axiom~A of Smale \cite{KH} if its nonwandering
set
$\Omega=\Omega(S)$ is hyperbolic with respect to~$S$ and the set of
periodic points of~$S$ is dense
in~$\Omega$. Recall also that the nonwandering set of a diffeomorphism
is closed and invariant under~$S$.
It is known that for Axiom~A dynamical systems the set
$\Omega$ is a finite (disjoint) union of closed, invariant and
transitive sets (i.e., there is a dense
orbit); each of these sets is called a basic set for~$S$.

By the continuity of the derivative of~$S$ and compactness (or
hyperbolicity), there are $C>0$ and
$\gamma>1$ so that
\[ d(S^n\theta,S^n\omega)\leq C\gamma^{|n|} d(\theta,\omega), \quad \
\forall\ \theta,\omega\in\Omega \ \ \mbox{and} \ \ \forall\
n\in\mathbb{Z}\ (\mbox{or} \ \mathbb{N}).
\] Now by taking $F:M\to\mathbb{R}$ continuously differentiable, the
Lipschitz
condition~(\ref{Lipschitz}) is immediately satisfied. So
Theorem~\ref{Transport} is applicable with $A$
being the set of initial conditions giving rise to periodic orbits
of~$S$. Therefore, there is a dense
$G_\delta$ set $\tilde \Omega\subset\Omega$ so that for each initial
condition $\omega\in\tilde\Omega$
the Schr\"odinger operator
$H^W_{\omega,S}$ presents quasi-ballistic dynamics.

It is interesting to note that due to hyperbolicity of~$\Omega$ the set
of periodic points of~$S$ is at
most countable, so that for ``chaotic'' Axiom~A systems the dense
$G_\delta$ set with quasi-ballistic dynamics is actually a nontrivial
one. 

Recall also that if $M$ is hyperbolic with respect to~$S$, then $S$ is
said to be an Anosov
diffeomorphism. These systems satisfy Axiom~A and so the above
conclusion about quasi-ballistic
dynamics holds. It seems to be an open question if for Anosov
diffeomorphisms the nonwandering sets
$\Omega$ always coincide with~$M$; in the case of Anosov diffeomorphism
on the torus $\mathbb{T}^2$ it is
known that there is just one basic set and it coincides with the whole
torus.

To the best of our knowledge, the only spectral specification related to such systems are a.s.\  purely
point spectrum (at the border of the spectrum) for hyperbolic toral automorphisms
$S$ on
$\mathbb{T}^2$ (i.e., a particular class of Anosov maps and so the torus $\mathbb{T}^2$ is a basic set)
and
$F\in C^1(\mathbb{T}^2)$ with zero average studied in
\cite{BoS}. Such spectral results are similar to those mentioned for the Doubling Map in
Subsection~\ref{appl.3}. Of course the periodic orbits of
$S$ generate absolutely continuous spectrum.

\

\subsection{Doubling Map}
\label{appl.3}
 Consider the operator $H^W_{\theta,S}$ defined by~(\ref{model}) on
$\ell^2(\mathbb{N})$ where $S$ is the transformation on $\Omega=[0,1]$
given by $S
\theta= 2 \theta$ (mod 1) and
$F=\cos:\Omega\to\mathbb{R}$. Note that $F$ satisfies the Lipschitz
condition. The set
$A=\{\theta$ whose expansion in the basis 2 is periodic$\}$ is dense in
$\Omega$. Since each element of $A$ corresponds to a periodic orbit
of~$S$, it follows from
\cite{La,dRJLS} that $A$ satisfies the hypotheses of
Theorem~\ref{Transport}. Furthermore, for every
$\theta,\omega\in\Omega=[0,1]$ and
$n\in\mathbb{N}$, one has
\[ d(S^n\theta,S^n\omega)=
|2^n\theta-2^n\omega|=2^{n}d(\theta,\omega).\]
 Therefore, by Theorem~\ref{Transport}, there exists a dense
$G_{\delta}$ set
$\tilde{\Omega} \subset [0,1]$ such that for any
$\theta\in\tilde{\Omega}$ and for every $p>0$,
$\beta^+_{\theta,S,W}(p)=1$---note that indeed this result holds for
any (nonconstant) periodic
continuously differentiable $F$.

Now fix (small) $\delta>0$ and $\lambda>0$ sufficiently small. Bourgain
and Schlag~\cite{BoS} has proven
that for $\ell-a.s. \ \theta\in [0,1]$, the operator $H_{\theta,S}^{0}$
has pure point spectrum in
$[-2+\delta,-\delta]\cup [\delta,2-\delta]$ with eigenfunctions
decaying exponentially. In particular,
$\Gamma^0_{S}(E)>0$ for
$E\in [-2+\delta,-\delta]\cup [\delta,2-\delta]$. Since in principle
$\tilde \Omega$ can have null
measure, we can not conclude that there are elements of $\tilde\Omega$
whose corresponding operator
has a point spectrum component.

\

\subsection{Chaotic One-Dimensional Maps}
\label{appl.Chaotic}
 Let $I$ be a compact interval in~$\mathbb{R}$ and $S:I\to I$ a
continuously differentiable map
(for simplicity we restrict ourselves to one-dimensional maps).
Suppose that restricted to
$\Lambda\subset I$ the map
$S$ is chaotic as defined by Devaney (\cite{Dev}, p.~50), i.e., it is
sensitive on initial
conditions, topologically transitive, and the periodic points are dense
in~$\Lambda$. The potential for
the Schr\"odinger operator~(\ref{model}) on $\ell^2(\mathbb{N})$ will
be the own orbits of~$S$, so that
$F$ is the identity map (or any other Lipschitz function). The hypotheses on
$F$ and $S$ in
Theorem~\ref{model} are clearly satisfied, as well as the existence of
the set~$A$.

Specific examples are the Tchebycheff polynomials \cite{Dev};
for
instance, $x\mapsto4x^3-3x$ and $x\mapsto8x^4-8x^2+1$ are chaotic
on~$[-1,1]$. For $r>2+\sqrt5$, the
map $S_r(x)=rx(1-x)$ is chaotic on the set
$\Lambda\subset[0,1]$ of points which never escape from $[0,1]$ upon
iterates of~$S_r$; for $r=4$ the map
 $S_4$ is chaotic on $\Lambda=[0,1]$.

Therefore, for the family of operators
$H^W_{x,S}$, with chaotic $S$ as above, there is a dense $G_\delta$ set
$\tilde{\Omega} \subset \Lambda$
such that for any
$x\in\tilde{\Omega}$ the corresponding operator presents
quasi-ballistic dynamics. It is a very
interesting open problem to say something about the spectra of such
operators. Are they ``in general'' pure
point as the intuition says? What about for $x\in\tilde{\Omega}$?

\

\subsection{Anderson Model}
\label{appl.4}
 Consider the operator $H^W_{\omega,S}$ defined by~(\ref{model}) on
$\ell^2(\mathbb{Z})$ where $S$ is the shift on
$\Omega=[-1,1]^{\mathbb{Z}}$ given by
$(S\omega)_j=\omega_{j+1}$ and $F:\Omega\to [-1,1]$ defined by
$F(\omega)=\omega_0$. Note that
$F(S^n\omega)=(S^n\omega)_0=\omega_n$. It is assumed that $\omega_n,\
n\in\mathbb{Z}$, are independent
identically distributed random variables with common probability measure
$\sigma$ not concentrated on a single point and $\int
|\omega_n|^{\alpha }d\sigma(\omega_n)<\infty$ for
some
$\alpha>0$. Denote by $\nu=\prod_{n\in\mathbb{Z}}\sigma$ the
probability measure on
$\Omega$. The metric on
$\Omega$ is given by
\[
d(\omega,\theta)=\sum_{j\in\mathbb{Z}}\frac{d_0(\omega_j,\theta_j)}{2^{|
 j|}},
\] where $d_0$ is the discrete metric. For every
$\omega,\theta\in\Omega$, one has
\[ |F(\omega)-F(\theta)|=|\omega_0-\theta_0|\leq 2d(\omega,\theta),
\] and so $F$ is Lipschitz. The set
$A$ of periodic sequences in $\Omega$ is dense in
$\Omega$. Since each periodic sequence determines a periodic orbit
of~$S$, it follows
from~\cite{La,dRJLS} that
$A$ satisfies the hypotheses of Theorem~\ref{Transport}. Furthermore,
for every
$\omega,\theta\in\Omega$ and for all $n\in\mathbb{Z}$, one has
\[ d(S^n\omega,S^n\theta)\leq 2^{|n|}\sum_{j\in\mathbb{Z}}
\frac{d_0(\omega_{j+n},\theta_{j+n})}{2^{|j+n|}} = d(\omega,\theta)\
2^{|n|}.
\] Therefore, by Theorem~\ref{Transport}, there exists a dense
$G_{\delta}$ set
$\tilde{\Omega} \subset \Omega$ such that for any
$\omega\in\tilde{\Omega}$ and for every $p>0$,\
$\beta^+_{\omega,S,W}(p)=1$.

In~\cite{CKM,SVW,vDK} it was proven that for $ \nu-a.s. \ \omega$,
$H^0_{\omega,S}$ has pure point spectrum with eigenfunctions decaying
exponentially. In particular,
$\Gamma^0_{S}(E)>0$ for every~$E$. As in
Application~\ref{appl.3}, we can not conclude that there are elements
of $\tilde\Omega$ whose
corresponding operators present point spectrum. Such quasi-ballistic
dynamics should be contrasted with
the dynamical localization proven
$\nu-a.s.$ \cite{GD} for this model.
In the sequel an important particular case is selected.

\

\subsubsection{\bf Bernoulli-Anderson Model} Take $H^W_{\omega,S}$ as
in Application \ref{appl.4} with
$\Omega=\{a_1,\ldots,a_k\}^{\mathbb{Z}},\ a_i\in\mathbb{R}$, and for
each
$n\in\mathbb{Z}$, $\sigma(\omega_n=a_i)=p_i,\ 0<p_i<1$ and
$\sum_{i=1}^k p_i=1$. The same conclusions of Application~\ref{appl.4}
hold.

\

\subsection{Rotations of $\mathbb{S}^1$ with analytic condition on $F$}
\label{appl.1}
 Consider the operator
$H^W_{(\theta,\alpha),S}$ defined by~(\ref{model}) with $S$ the
transformation on
$\Omega_a:=\mathbb{S}^1\times [-a,a]$, $a>0$ fixed, given by
$S(\theta,\alpha)=(\theta+\pi\alpha,\alpha)$, and $F=g\circ\pi_1$, with
$g:\mathbb{S}^1\to\mathbb{R}$ nonconstant analytic of period~$1$ and
$\pi_1:\Omega_a\to\mathbb{S}^1$ the projection
$\pi_1(\theta,\alpha)=\theta$. For every
$(\theta,\alpha),(\omega,\beta)\in\Omega_a$, it follows by the Mean
Value Theorem that
\begin{eqnarray*}\left|F(\theta,\alpha)-F(\omega,\beta)\right| & = &
\left|g(\theta)-g(\omega)\right|
\\ &\leq & \left(\ \sup_{z\in\mathbb{S}^1}|g'(z)|\right)|\theta-\omega|
\\ &\leq &\ L\ d((\theta,\alpha),(\omega,\beta)),
\end{eqnarray*} where $L= \sup_{z\in\mathbb{S}^1}|g'(z)|$ and
$d((\theta,\alpha),(\omega,\beta))=\sqrt{(\theta-\omega)^2 +
(\alpha-\beta)^2}$; in other words,
$F$ satisfies the Lipschitz condition. The set
\[ A=\{(\theta,\alpha_0):
\theta\in\mathbb{S}^1, \alpha_0\in
\mathbb{Q} \cap [-a,a] \}
\]
 is dense in $\Omega_a$ and for each
$(\theta,\alpha_0)\in A$, \ $S^n(\theta,\alpha_0)$ describes a periodic
orbit at the ``height''
$\alpha_0$. Therefore the potential $\lambda F(S^n(\theta,\alpha_0))$
is periodic and, due
to~\cite{La,dRJLS},
$A$ satisfies the hypothesis (\ref{rel2Theorem1}) of
Theorem~\ref{Transport}. Now note that
 for every
$(\theta,\alpha),(\omega,\beta)\in\Omega_a$ and
$n\in\mathbb{Z}\setminus\{0\}$ ($n=0$ is trivial), one has
\begin{eqnarray*} d(S^n(\theta,\alpha),S^n(\omega,\beta)) &\leq&
d(S^n(\theta,\alpha),S^n(\theta,\beta))+
d(S^n(\theta,\beta),S^n(\omega,\beta)) \\ &=&
\sqrt{n^2\pi^2(\alpha-\beta)^2+(\alpha-\beta)^2}+
\sqrt{(\theta-\omega)^2} \\ &\leq& \left(\sqrt{\pi^2+1}\
|\alpha-\beta|+|\theta-\omega|\right)|n| \\
&\leq&
\left(\sqrt{\pi^2+1}+1\right)\ d((\theta,\alpha),(\omega,\beta))\ |n|.
\end{eqnarray*} Therefore, by Theorem~\ref{Transport}, there exists a
dense
$G_{\delta}$ set $\tilde{\Omega}_a \subset\Omega_a$ such that for any
$(\theta,\alpha)\in\tilde{\Omega}_a$ and for every $p>0$,
$\beta^+_{(\theta,\alpha),S,W}(p)=1$, with $W$
satisfying~(\ref{perturbation}). Observe that
$F(S^n(\theta,\alpha))=g(\theta+n\pi\alpha)$.

It follows by Sorets and Spencer~\cite{SS} that there exists a number
$\lambda_0(F)>0$ such that for
$\lambda>\lambda_0$, \ $\Gamma^0_{S}(E)>0$ for every~$E$, every
irrational $\alpha$ and
$\ell-a.s. \ \theta$ ($\ell$ on
$\mathbb{S}^1$ is ergodic with respect to $\pi_1\circ S$). Since the
generic set $\tilde \Omega_a$ can have
zero measure, we are not assured to be able to apply Theorem~\ref{Point
Spectrum} to elements of
$\tilde \Omega_a$ in order to obtain quasi-ballistic dynamics with pure
point spectrum. Nevertheless, the
original view in~\cite{dRJLS,GKT} (i.e., to consider for each $\alpha$
a different map) for the cosine
function, implies uniformity in~$\theta$ also in our case, so that we
get new examples of Schr\"odinger
operators with pure point spectrum and quasi-ballistic dynamics. Let us
reconsider the construction, since
it will also be employed in Application~\ref{appl.2}.

Begin by replacing (rewriting, in fact) $H^W_{(\theta,\alpha),S}$ with
$H^W_{\theta,S_\alpha}$, where
$S_\alpha(\theta)=\theta+\pi\alpha$, $\theta \in \mathbb{S}^1$, and
$F=g$ (take $\pi_1$ as the identity).
For each $\alpha_0\in\mathbb{Q}$ and $\theta \in \mathbb{S}^1$, the
potential $\lambda
F(S_{\alpha_0}^n(\theta))$ is periodic and there is
$J^\theta_{\alpha_0}\subset
\sigma(H^0_{\theta,S_{\alpha_0}})$ with $\ell(J^\theta_{\alpha_0})>0$
so that uniformly in $\theta$
\[
\|\Phi^0_{\theta,S_{\alpha_0}}(E,n,0)\|\le C_{\alpha_0},\quad \forall
E\in J^\theta_{\alpha_0},n\in \mathbb{Z}.
\] Furthermore, for all $\theta$ and $n$
\[ d(S^n_\alpha(\theta),S^n_{\alpha_0}(\omega)) =
d(\theta+n\pi\alpha,\theta+n\pi\alpha_0)= \pi
|\alpha-\alpha_0|\, |n|.
\] By repeating the arguments of Lemma~\ref{q-b.motion} and
Theorem~\ref{Transport}, but now with
\[
\tilde{B_n}=\left\{ \alpha \in [-a,a]\Big|\ \exists\
\epsilon<\frac{1}{n} : \forall
\theta\in\mathbb{S}^1,
\ K_{\mu^W_{\theta,S_\alpha}}(q,\epsilon)\geq \ C_q \
\frac{\epsilon^{-1+q}}{\log(\epsilon^{-1})} \right\}
\] instead of $B_n$, one concludes that there exists a dense $G_\delta$
set of irrational numbers
$\mathcal G\subset[-a,a]$, so that for each fixed $\alpha\in\mathcal G$
and every $\theta\in\mathbb{S}^1$,
the operator
$H^W_{(\theta,\alpha),S}=H^W_{\theta,S_\alpha}$ presents
quasi-ballistic dynamics.

Therefore, by Theorem~\ref{Point Spectrum}, for $\alpha \in\mathcal G$
and $\ell-a.s. \ \theta$ the
operator
$H^W_{(\theta,\alpha),S}$ with nonconstant analytic $F$ on the half
lattice $\ell^2(\mathbb{N})$
case, under the rank one perturbation
$W=\kappa\langle\delta_1,
\cdot\rangle\delta_1$,
$\lambda>\lambda_0$ and $\alpha$ irrational, has pure point spectrum for
$\ell-a.s. \ \kappa,$ and also presents quasi-ballistic dynamics.

\

 Now some interesting particular cases of potentials generated by this
dynamical system will be described.

\

\subsubsection{\bf Almost Mathieu}\label{AlmostMathieu} This is just a
reconsideration of the
``pathological example'' of~\cite{dRJLS}.
$H^W_{(\theta,\alpha),S}$ is defined by~(\ref{model}), where
$S$ is the transformation on
$\Omega_a$ given by
$S(\theta,\alpha)=(\theta+\pi\alpha,\alpha)$,
$F=\cos\circ\pi_1:\Omega_a\to\mathbb{R}$, $W=\kappa\langle\delta_1,
\cdot\rangle\delta_1$, $\alpha$ is irrational and
$\lambda>\lambda_0=2$. Under such conditions both
Theorems~\ref{Transport} and~\ref{Point Spectrum} hold
for proper sets, as discussed in Application~\ref{appl.1}.

\

\subsubsection{\bf Circular Billiards~\cite{ChM}} The potential now is
along the orbits of a particle
under specular reflections on a circular billiard.
$H^W_{(r,\phi),S}$ is defined by~(\ref{model}), where
$S$ is the transformation on $\Omega_{\frac{\pi}{2}}$ given by
$S(r,\phi)=(r+\pi-2\phi,\phi)$ and $F=g\circ\pi_1$ with
$g:\mathbb{S}^1\to\mathbb{R}$ nonconstant analytic of period~1. Again
the conclusions of
Application~\ref{appl.1} hold.

\

\subsubsection{\bf Twist Map \cite{KH}}
$H^W_{(\theta,r),S}$ is defined by~(\ref{model}), with
$S$ the transformation on
$\Omega=\overline{D(0,1)}$ (closed disk of center~0 and radius~1
in~$\mathbb{R}^2$) given by
$S(\theta,r)=(\theta+\rho(r),r),$ with
$\rho:[0,1]\to[0,2\pi]$ continuous, $\rho(0)=0,\ \rho'(r)>0$, and
$F=g\circ\pi_1$ with $g$ a nonconstant real-analytic function of
period~1. So, the potential is defined
along orbits of an integrable twist map, and Theorems~\ref{Transport}
and~\ref{Point Spectrum} hold concomitant for
proper sets.

\

\subsection{Rotations of the torus with analytic condition on $F$}
\label{appl.2}
 Consider the operator $H^W_{(\theta,\alpha),S}$ defined
by~(\ref{model}), where $S$ is the
transformation on
$\Omega_a^k:=\mathbb{T}^k\times [-a,a]^k$ ($\mathbb{T}^k$ is the
k-dimensional torus;
$a>0$ fixed) given by
$S(\theta,\alpha)=(\theta+\pi\alpha,\alpha)$, with
$\theta=(\theta_1,\ldots,\theta_k),\
\alpha=(\alpha_1,\ldots,\alpha_k)$,
$F=g\circ\pi_k$ with
$g:\mathbb{T}^k\to\mathbb{R}$ nonconstant analytic of period~$1$ in
each component, and
$\pi_k:\Omega_a^k\to\mathbb{T}^k$ the projection
$\pi_k(\theta,\alpha)=\theta$. Observe that for $k=1$, we are in the
case of Application~\ref{appl.1}
above. Similarly to \ref{appl.1}, one obtains a dense
$G_{\delta}$ set $\tilde{\Omega}_a^k \subset\Omega_a^k$ such that for
any
$(\theta,\alpha)\in\tilde{\Omega}_a^k$ and for every
$p>0$, $\beta^+_{(\theta,\alpha),S,W}(p)=1$. Note that
$F(S^n(\theta_1,\ldots,\theta_k,\alpha_1,\ldots,\alpha_k))=
g(\theta_1+n\pi\alpha_1,\ldots,\theta_k+n\pi\alpha_k)$.

The construction of $\mathcal G$ in Application~\ref{appl.1} has a
direct counterpart here, so that
$\beta^+_{(\theta,\alpha),S,W}(p)=1$ for $\alpha$ in a dense $G_\delta$ set
$\mathcal G_k\subset[-a,a]^k$ and
every $\theta$. The above mentioned result of Sorets and Spencer is
still valid in this
case~\cite{Bourg}: there is  $\lambda_0>0$ so that if
$\lambda>\lambda_0$, then
$\Gamma^0_{S}(E)>0$ for every~$E$, every
 incommensurate vector $\alpha$ (i.e., $\alpha\cdot j\ne0$ for all
$j\in\mathbb{Z}^k\setminus\{0\}$) and
$\ell-a.s.$
$\theta=(\theta_1,\ldots,\theta_k)$ ($\ell$ on
$\mathbb{T}^k$ is ergodic with respect to $\pi_k\circ S$). Therefore,
by Theorem~\ref{Point Spectrum},
for such $\alpha$'s the operator
$H^W_{(\theta,\alpha),S}$ on $\ell^2(\mathbb{N})$, with
$W=\kappa\langle\delta_1,
\cdot\rangle\delta_1$ and $\lambda$ large enough, has pure point
spectrum for
 $\ell-a.s. \ \theta$ and $\kappa$. Since necessarily a $G_\delta$ set
in $[-a,a]^k$ contains
incommensurate vectors
$\alpha$ (in particular for $\mathcal G_k$), again we have got new
examples of Schr\"odinger operators
with pure point spectrum and quasi-ballistic dynamics. We stress once
more that, in fact, our arguments
come from a (simple) closer inspection of the original arguments of
\cite{dRJLS,GKT} for the
Almost-Mathieu operator.

\

\section{The Discrete Dirac Model}
\label{DiracSection} The single particle one-dimensional discrete Dirac
operator was studied
in~\cite{dOP1,dOP2} and is described by $(\omega\in\Omega)$
\[
\textbf{D}_{\omega}(m,c):=\left(
\begin{array}{cc} mc^2 & cD^* \\ cD & -mc^2
\end{array} \right) + V_{\omega}Id_2,
\] acting on $\ell^2(\mathbb{Z},\mathbb{C}^2)$ or
$\ell^2(\mathbb{N},\mathbb{C}^2)$, where $Id_2$ is the $2\times2$ identity
matrix, $c>0$ represents the
speed of  light,
$m\geq 0$ is the mass of the particle, $D$ is the finite difference
operator defined by
$(D\psi)(n)=\psi(n+1)-\psi(n)$ and $D^*$ is the adjoint of~$D$.

Besides being a physical model, it was of interest because for the
massless case and two-valued Bernoulli
potentials, its behavior is similar to the corresponding Schr\"odinger
case after dimerization, with
presence of the so-called critical energies
\cite{JSBS,dOP2}. So, under certain conditions it is possible to get
pure point spectrum and nontrivial
transport a.s.\ with
$\beta^-(p)\ge (1-\frac{1}{2p})$.

 By considering the potential
$V_{\omega}(n)=\lambda F(S^n\omega)+W(n)$, with $F$ and $W$ satisfying
(\ref{Lipschitz}) and
(\ref{perturbation}) respectively, Theorems~\ref{Transport}
and~\ref{Point Spectrum} hold, and so all
applications in Section~\ref{ApplicationsSection} have a counterpart
for this model; this follows after a
huge set of technical details (not presented) are checked and adapted
by following the lines
of~\cite{dOP2}. Hence this discrete Dirac version, with suitable
potentials along some dynamical systems,
provides examples of relativistic quantum operators with
quasi-ballistic dynamics, some also with point
spectrum.

\

\end{document}